\documentclass[onecolumn,amssymb,preprintnumbers,prc,floatfix]{revtex4-1}
\usepackage{rotating,epsfig,dcolumn}
\usepackage{graphicx,bm}

\begin{document}
\title{Novel Nuclear Structure Aspects of the 0$\nu$$\beta\beta$-decay}

\author{ J. Men\'endez$^{*,a,b}$, A. Poves$^*$, E. Caurier$^{\dagger}$ and F. Nowacki$^{\dagger}$}

\affiliation{$^*$Departamento de F\'{\i}sica Te\'orica, 
  and IFT, UAM-CSIC, Universidad Aut\'onoma  de Madrid, 28049-Madrid, Spain}
\affiliation{$^a$Institut f\"ur Kernphysik, Technische Universit\"at
Darmstadt, 64289 Darmstadt, Germany}
\affiliation{$^b$ExtreMe Matter Institute EMMI, GSI Helmholtzzentrum f\"ur
Schwerionenforschung GmbH, 64291 Darmstadt, Germany}
\affiliation{$^{\dagger}$IPHC, IN2P3-CNRS/Universit\'e Louis Pasteur, 67037-Strasbourg France.}

\begin{abstract} 
We explore  the influence of the deformation on the nuclear matrix elements of the neutrinoless double beta
decay (NME), concluding that the difference in deformation -or more generally in the amount of quadrupole
correlations- between parent and grand daughter nuclei  quenches strongly the decay.  We correlate these
differences with the seniority structure of the nuclear wave functions. In this context, we examine
the present discrepancies between the NME's obtained in the framework of the Interacting Shell Model 
and the Quasiparticle RPA. In our view,  part of the discrepancy can be due to the limitations of the
spherical QRPA in treating nuclei which have strong quadrupole correlations. We surmise that the NME's
in a basis of generalized seniority are approximately model independent, $i. \; e.$ they are "universal".

\end{abstract}
\maketitle

\section{Introduction}
\label{intro}

The double beta decay is a rare weak process which takes place between
two even-even isobars when the single beta decay is energetically
forbidden or hindered by large spin difference. The two neutrino beta
decay is a second order weak process ---the reason of its low rate---,
and has been measured in a few nuclei. The $0\nu\beta\beta$ decay
is analog but requires neutrinos to be Majorana fermions. With the exception
of one unconfirmed claim \cite{KlapdorKleingrothaus:2004wj},
it has never been observed, and currently there is a number of experiments
either taking place  or expected for the near future ---see e.g. ref. \cite{Avignone:2007fu}---
devoted to detect this process and to set up firmly the nature of
neutrinos.
Furthermore, the $0\nu\beta\beta$ decay is also sensitive to the absolute
scale of the neutrino mass, and hence to the mass hierarchy. Since
the half-life of the decay is determined, together with the masses,
by the nuclear matrix element for the process, its knowledge
is essential to predict the most favorable decays and,
once detection is achieved, to settle the neutrino mass scale and
hierarchy.

Two different  methods were traditionally used to calculate the
NME's for $0\nu\beta\beta$ decays, the quasiparticle
random-phase approximation and the shell model in large valence spaces
(ISM). The QRPA 
has produced   results for most
of the possible emitters since long \cite{Kortelainen:2007rh,Kortelainen:2007mn,Rodin:2007fz}. 
The ISM, that was limited to  a few cases till recently \cite{Caurier:2007wq},  can nowadays
describe (or will do it shortly)  all  the experimentally relevant decays but one, the decay of 
$^{150}$Nd.  Other
approaches, that share a common prescription for the transition operator
 (including higher order corrections), and for the treatment
of the short range correlations (SRC) and  the finite size effects, are the Interacting Boson Model
\cite{Barea:2009ia}, and the Projected Hartree Fock Bogolyuvov method \cite{Chandra:2009my}.

The expression for the half-life of the
$0\nu\beta\beta$ decay can be written as \cite{Doi:1985dx}:

\begin{equation}
\left(T_{1/2}^{0\nu\beta\beta}\left(0^{+}\rightarrow0^{+}\right)\right)^{-1}=G_{01}\left|M^{0\nu\beta\beta}\right|^{2}\left(\frac{\left\langle m_{\beta\beta}\right\rangle }{m_{e}}\right)^{2},\label{eq:t-1}\end{equation}

\noindent
where $\left\langle m_{\beta\beta}\right\rangle =\left|\sum_{k}U_{ek}^{2}m_{k}\right|$
is the effective neutrino mass, a combination of the neutrino mass eigenvalues $m_{k}$. 
 $U$ is the neutrino mixing matrix and $G_{01}$
is a kinematic factor dependent on the charge, mass and available
energy of the process. $M^{0\nu\beta\beta}$ is the nuclear matrix element of the neutrinoless
double beta decay operator, which has Fermi, Gamow-Teller and Tensor components. 
The kinematic factor $G_{01}$ depends on the value of the coupling constant $g_{A}$. In addition, some
calculations use different values of r$_0$ in the formula R=r$_0$ A$^{1/3}$.
It is therefore convenient to define:

\begin{equation}
M'^{\;0\nu\beta\beta}=\left(\frac{g_{A}}{1.25}\right)^{2} \left(\frac{1.2}{r_0}\right)M^{0\nu\beta\beta}
\end{equation}

 In this way the theoretical $M'^{\;0\nu\beta\beta}$'s are directly comparable among them 
irrespective of  the values of $g_{A}$ and r$_0$ employed in their calculation, since they share a common
$G_{01}$ factor ---the one computed with  $g_{A}=1.25$ and r$_0$=1.2~fm. Thus, the translation of the
$M'^{\;0\nu\beta\beta}$'s into half-lives is transparent.

\section{Pairing and Quadrupole; The Influence of Deformation}
\label{sec:1}

An important issue regarding the $0\nu\beta\beta$ decay is the role of the correlations; 
pairing that drives the nucleus toward a superfluid state and quadrupole that favors deformed
intrinsic shapes. It has been show recently that  the $2\nu\beta\beta$ is hindered by
the difference in deformation  between the initial and final nuclei
\cite{Simkovic:2004pf,AlvarezRodriguez:2004uk}. For the neutrinoless mode,
the calculations 
\cite{Caurier:2007wq} indicate that the pairing interaction favors the 
decay and that, consequently, the truncations in seniority, which quench
the pair breaking action of the quadrupole correlations, produce an
overestimation of  the values of the NME's. On the other hand, the NME's are also
reduced when the parent and grand-daughter nuclei have different deformations
\cite{Caurier:2007qn,Simkovic:2007me}. 

\begin{center}
\begin{figure*}[h]
    \leavevmode
\resizebox{0.7\textwidth}{!}{%
    \includegraphics[angle=0]{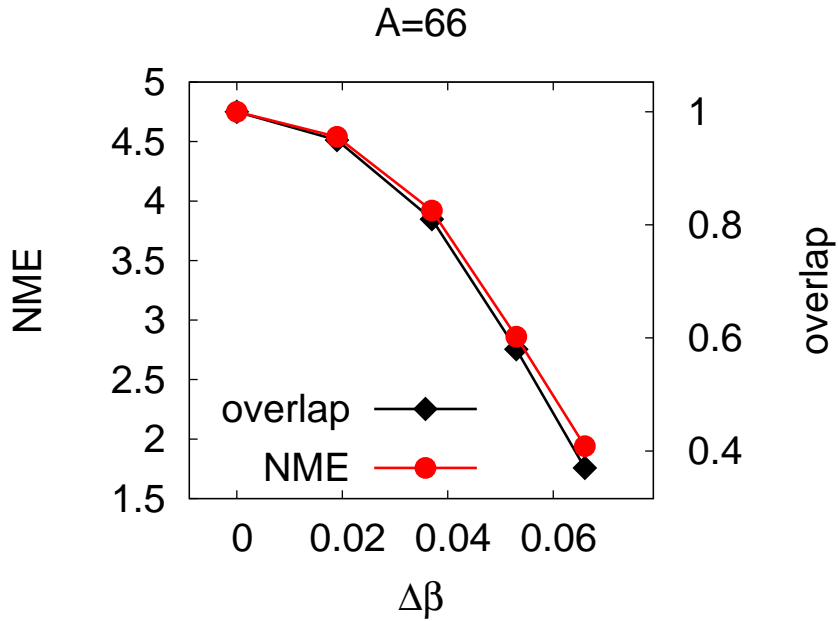}}    
  \caption{$^{66}$Ge $\rightarrow$ $^{66}$Se
NME, $M'^{0\nu}$, as a function of the difference in deformation
induced by  the extra
quadrupole interaction added to  $^{66}$Se.
}
  \label{fig:a66}
  \end{figure*}
\end{center}

We have chosen to  study the (unphysical) transition between the mirror nuclei  $^{66}$Ge and  $^{66}$Se
in order to have a clearer view of the effect of the deformation in the NME's. This 
transition has the peculiarity  that the wave functions of the initial
and final nuclei are identical (provided Coulomb effects are neglected)
and consequently it is easier to disentangle the contributions of the $0\nu\beta\beta$ operator and  the nuclear 
wave functions  to the NME.  The calculations are carried out in the valence space  r3g with the effective interaction gcn28:50. 
The SRC are modeled by a Jastrow factor with
the Spencer and Miller parametrization \cite{Miller:1975hu}, although it has been shown recently that,
once the finite size of the nucleon has been taken into account by a dipole
form factor, softer options are more realistic \cite{Engel:2009ha,Simkovic:2009pp} 

 To increase the deformation of a given nucleus we add to the
effective interaction a term $\lambda \; Q\cdot Q$.
Fig. \ref{fig:a66} shows the results when the final nucleus has been
artificially deformed by adding an extra quadrupole-quadrupole term.
Notice in the first place that  for $\lambda$=0  both nuclei are deformed
with $\beta$~$\sim$~0.2. In spite of that, the NME is a factor of two larger
than the  values obtained for the A=76 and A=82 decays in the
same valence space and with the same interaction. Hence, even if the
two A=66 partners are deformed, the fact that their wave functions are identical
enhances the decay.  Nevertheless,  the NME is still far from its  expected value
in the superfluid limit (NME$\sim$8).   
The figure shows that the reduction of the NME as the difference in deformation
increases is very pronounced.
For the values of $\lambda$ between 0.0 and 0.2, the difference
in deformation parameter between parent and grand daughter grows from
zero to about 0.1.
In addition, the NME  follows closely the overlap between the  wave function of
one nucleus obtained with $\lambda$=0 and the wave function of the same nucleus
obtained with $\lambda$$\ne$0.
This means that, if we write the final wave function as:
 $\left|\Psi_{\,}\right\rangle =a\left|\Psi_{0}\right\rangle +b\left|\Psi_{qq}\right\rangle $,
the $0\nu\beta\beta$ operator does not connect $\Psi_{0}$ and   $\Psi_{qq}$.
This behavior of the NME's  with respect to the difference of deformation between parent and grand daughter  
is common to all the  transitions between mirror nuclei that we have studied (A=50, A=110) and to more
realistic cases like
the A=82 decay that we have examined in detail in \cite{Menendez:2009def}. Therefore we can submit that
this is a robust result. Similar results hold also for the 2$\nu$ decays.

\section{The NME's and the seniority structure of the nuclear wave functions}

 We can also analyze the results of the preceding section in terms of the seniority 
 structure of the wave functions of parent and grand
 daughter nuclei. Indeed when $\Delta \beta$=0 both  $^{66}$Ge and  $^{66}$Se have identical wave functions.
 The probabilities of the components of different seniority are given in  table \ref{tab:a66}.
  It is seen that changing  $\beta$ from
 0.22 (mildly deformed) to 0.30 (strongly deformed) increases drastically the amount of high seniority components
 in the wave function, provoking a seniority mismatch between the decaying and the final nuclei. This leads
 to very  large cancelations of the nuclear matrix elements of the decay, as shown also in table \ref{tab:a66}.

 \begin{center}
\begin{table}[h]
\caption{{\label{tab:a66}} The seniority structure of the wave functions in the A=66 mirror decay}
\begin{center}
\begin{tabular*}{0.9\textwidth}{@{\extracolsep{\fill}}l|ccccc}
\hline \hline
 & $s=0$ & $s=4$ & $s=6$ & $s=8$ & $s=10$ \\
\hline
  $\Delta \beta$=0 & 39 & 43 &  7 &  10 & 1 \\
 $\Delta \beta$=0.08 & 6 & 32 & 21 & 31 & 10 \\
\hline
  & M$^{0\nu}_F$  &   M$^{0\nu}_{GT}$    & M$^{0\nu}_T$  & \multicolumn{2}{l}{M'$^{0\nu}$} \\
\hline
 $\Delta \beta$=0          & -2.02 & 3.95 &  0.08 &  5.16 \\
 $\Delta \beta$=0.08   & -0.76 & 1.65 &  0.02 &  2.12 \\
\hline \hline
\end{tabular*}
\end{center}
\end{table}
\end{center}

\begin{figure*}[h]
\begin{center}
\resizebox{1.0\textwidth}{!}{%
  \includegraphics[angle=0]{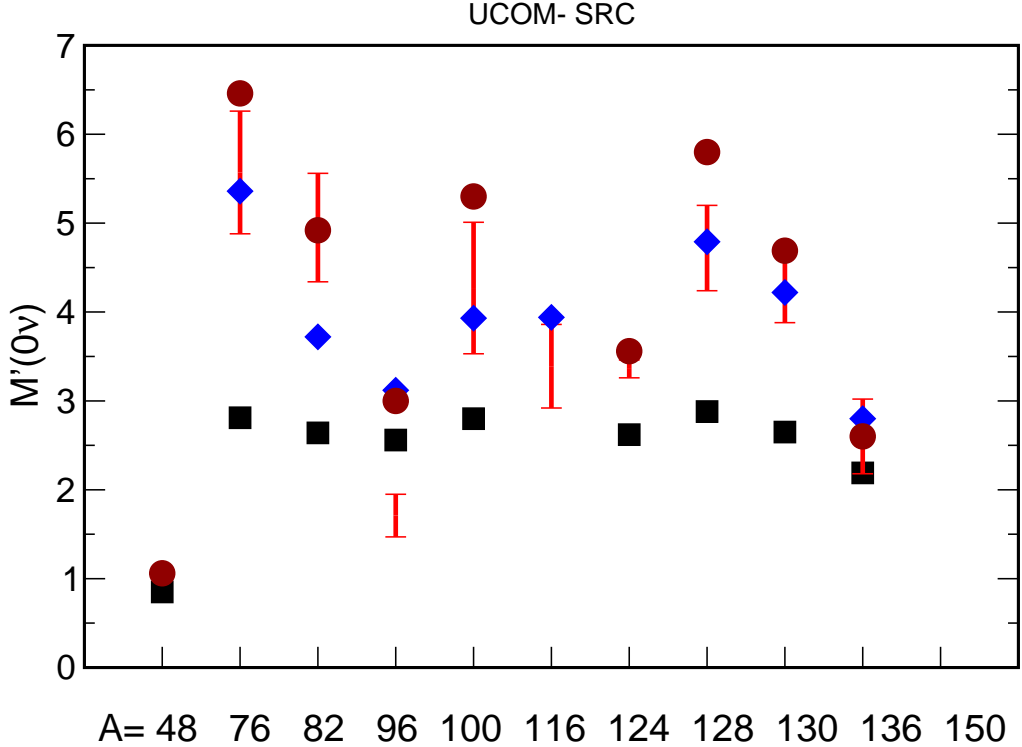}}
\caption{The neutrinoless double beta decay nuclear matrix elements $M'^{\;0\nu\beta\beta}$ for ISM and QRPA 
calculations treating the SRC with the UCOM approach. Tu, QRPA results from
ref. \protect\cite{Simkovic:2007vu} and Jy, QRPA results from refs. \protect\cite{Kortelainen:2007rh,Kortelainen:2007mn}.
The ISM results for A=96 and A=100 are preliminary}
\label{fig:ucom}       
\end{center}
\end{figure*}

Coming back to the physically relevant decays, we compare in figure \ref{fig:ucom} 
 the ISM and QRPA  NME's. In both approaches, the
SRC are taken into account in the UCOM framework  \cite{Feldmeier:1997zh} and g$_A$=1.25
is adopted. We have discussed elsewhere that the discrepancies between both approaches
show the following trends:  when the nuclei that participate in the decay have  a low
level of quadrupole correlations, as in the decays of $^{96}$Zr,   $^{124}$Sn and  $^{136}$Xe,
the calculations tend to agree.  On the contrary, when the correlations are large, the QRPA 
in a spherical basis seems not to be able to capture them fully.  As the effect of the correlations
is  to reduce the NME's, the QRPA produces NME's that are too large in
$^{76}$Ge, $^{82}$Se, $^{100}$Mo, $^{128}$Te, and $^{130}$Te.  Indeed, when the ISM calculations
are truncated to maximum seniority s$_m$=4, which is the leading order of the ground state correlations
in the QRPA (corresponding to the two quasi-particle contribution), they follow closely the QRPA results,
as can be seen  also in figure \ref{fig:ucom}. Notice that only when the ISM calculations are converged at this level
of truncation the two approaches do produce similar NME's.
\begin{center}
\begin{table}[h]
\caption{{\label{tab:nu0nu}} The seniority structure of the wave functions in the ISM and QRPA}
\begin{center}
\begin{tabular*}{0.9\textwidth}{@{\extracolsep{\fill}}c|cccccccc}
\hline \hline
 & $s=0$ & $s=4$ & $s=6$ & $s=8$ & $s=10$ & $s=12$ & $s=14$ & $s=16$\\
\hline
  & \multicolumn{8}{c}{ISM} \\
\hline
$^{48}$Ca & 97 & 3 &  - &  - & - & - & - & -\\
$^{48}$Ti & 59 & 36 &  4 &  1 & - & - & - & -\\
$^{76}$Ge & 43 & 41 &  7 &  8 & 1 & - & - & -\\
$^{76}$Se & 26 & 41 & 11 & 16 & 4 & 1 & - & -\\
$^{82}$Se & 50 & 39 & 10 &  1 & - & - & - & -\\
$^{82}$Kr & 44 & 41 &  6 &  8 & 1 & - & - & -\\
$^{128}$Te & 70 & 26 & 3 &  1 & - & - & -& -\\
$^{128}$Xe & 37 & 41 & 9 & 10 & 2 & - & - & -\\
\hline
  & \multicolumn{8}{c}{QRPA} \\
\hline
$^{76}$Ge & 55 & 33 &  - &  10 & - & 2 & - & -\\
$^{76}$Se & 59 & 31 & - & 8 & - & 2 & - & -\\
$^{82}$Se & 56 & 32 & - &  9 & - & 2 & - & -\\
$^{82}$Kr & 54 & 34 &  - &  11 & - & 2 & - & -\\
$^{128}$Te & 52 & 34 & - &  11 & - & 3 & -&- \\
$^{128}$Xe & 40 & 37 & - & 17 & -  & 5 & - & 1\\
\hline \hline
\end{tabular*}
\end{center}
\end{table}
\end{center}

 We compare in table \ref{tab:nu0nu} the seniority structure of the wave functions of the ISM and QRPA,
 in some of the cases for which  the latter are available \cite{Escuderos:2010xx}. It is seen that the differences are important
 and share a common trend:  in the QRPA, the seniority structure of parents and grand daughters is much
 more similar than in the ISM.  According to what we have seen in the A=66 case, this is bound to produce
 larger NME's in the QRPA than in the ISM, as it is actually the case. To make this statement quantitative, we
 have  developed the ISM matrix elements in a basis of generalized seniority
 
 $$   M_{F,GT,T} =  \sum_{\alpha,\beta}    A_{\nu_i(\alpha)}   B_{\nu_f(\beta)} 
 \langle \nu_f(\beta)  | O_{F,GT,T} | \nu_i(\alpha) \rangle $$

\noindent
where the A's and B's are the amplitudes of the different seniority components of the wave functions 
of the initial and final nuclei. Obviously, when we plug the ISM amplitudes in this formula, we recover
the ISM NME's. But, what shall we obtain if we  put the QRPA amplitudes instead? Indeed, we get 
approximately the QRPA NME's!  (5.73 for A=76 and 4.15 for A=82). Therefore as we had anticipated,
the seniority mismatch of the initial and final wave functions, which is severely underestimated in the
QRPA calculations, explains most of the discrepancy between the two descriptions.   In addition, this
result strongly suggests that there is some kind of universal behavior in the NME's of the neutrinoless double
beta decay when they are computed in a basis of generalized seniority.  If this is so, the only relevant difference
between the different theoretical approaches would reside in the seniority structure of the wave functions that they produce. 

\begin{center}
\begin{table}[h]
\caption{{\label{tab:nume}} The GT NME's of the A=48 decay in the generalized seniority basis}
\begin{center}
\begin{tabular*}{0.9\textwidth}{@{\extracolsep{\fill}}c|cccc}
\hline \hline
$^{48}$Ti  & $s=0$ & $s=4$ & $s=6$ & $s=8$ \\
\hline
$^{48}$Ca $s=0$ & 3.95 & -3.68 &  - &  - \\
$^{48}$Ca $s=4$ & 0.00 & -0.26 &  0.08 &  -0.02 \\
\hline 
\end{tabular*}
\end{center}
\end{table}
\end{center}

A very spectacular example of the cancellation of the NME by the seniority mismatch is provided by the $^{48}$Ca
decay. In Table \ref{tab:nu0nu}  we have included also the seniority structures of the two nuclei, and we see that they
are very different. If we now examine the values of the matrix elements
$\langle \nu_f(\beta)  | O_{GT} | \nu_i(\alpha) \rangle$ we find the values listed in Table \ref{tab:nume}. There are
two large matrix elements one diagonal and another off-diagonal of the same size and opposite sign. If the two nuclei
were dominated by the seniority zero components one should 
obtain  M$_{GT}$$\sim$4. If   $^{48}$Ti were a bit more deformed, M$_{GT}$ will be essentially zero. The value
produced by the KB3 interaction is 0.75 that is more than a factor five reduction with respect to the seniority zero
limit.  Earlier work on double beta decays in a basis of generalized seniority (limited to s=0 and s=4 components)
showing also this kind of cancellations can be found in ref.~\cite{Engel:1989vj} 

\section*{Acknowledgments}
  This work is partly supported by the Spanish Ministry of
 Ciencia e Innovaci\'on under grant FPA2009-13377, by the 
 Comunidad de Madrid (Spain) project HEPHACOS S2009/ESP-1473,
by the IN2P3(France)-CICyT(Spain) collaboration agreements, by
the DFG through  grant SFB 634 [J.M.] and by
the Helmholtz Association through the Helmholtz Alliance Program,
contract HA216/EMMI ``Extremes of Density and Temperature: Cosmic Matter
in the Laboratory'' [J.M.].

\bibliography{d_beta_s}

\begin{thebibliography}{21}%
\makeatletter
\providecommand \@ifxundefined [1]{%
 \@ifx{#1\undefined}
}%
\providecommand \@ifnum [1]{%
 \ifnum #1\expandafter \@firstoftwo
 \else \expandafter \@secondoftwo
 \fi
}%
\providecommand \@ifx [1]{%
 \ifx #1\expandafter \@firstoftwo
 \else \expandafter \@secondoftwo
 \fi
}%
\providecommand \natexlab [1]{#1}%
\providecommand \enquote  [1]{``#1''}%
\providecommand \bibnamefont  [1]{#1}%
\providecommand \bibfnamefont [1]{#1}%
\providecommand \citenamefont [1]{#1}%
\providecommand \href@noop [0]{\@secondoftwo}%
\providecommand \href [0]{\begingroup \@sanitize@url \@href}%
\providecommand \@href[1]{\@@startlink{#1}\@@href}%
\providecommand \@@href[1]{\endgroup#1\@@endlink}%
\providecommand \@sanitize@url [0]{\catcode `\\12\catcode `\$12\catcode
  `\&12\catcode `\#12\catcode `\^12\catcode `\_12\catcode `\%12\relax}%
\providecommand \@@startlink[1]{}%
\providecommand \@@endlink[0]{}%
\providecommand \url  [0]{\begingroup\@sanitize@url \@url }%
\providecommand \@url [1]{\endgroup\@href {#1}{\urlprefix }}%
\providecommand \urlprefix  [0]{URL }%
\providecommand \Eprint [0]{\href }%
\@ifxundefined \urlstyle {%
  \providecommand \doi  [0]{\begingroup \@sanitize@url \@doi}%
  \providecommand \@doi [1]{\endgroup \@@startlink {\doibase
  #1}doi:\discretionary {}{}{}#1\@@endlink }%
}{%
  \providecommand \doi  [0]{doi:\discretionary{}{}{}\begingroup
  \urlstyle{rm}\Url }%
}%
\providecommand \doibase [0]{http://dx.doi.org/}%
\providecommand \Doi [0]{\begingroup \@sanitize@url \@Doi }%
\providecommand \@Doi  [1]{\endgroup\@@startlink{\doibase#1}\@@Doi}%
\providecommand \@@Doi [1]{#1\@@endlink}%
\providecommand \selectlanguage [0]{\@gobble}%
\providecommand \bibinfo  [0]{\@secondoftwo}%
\providecommand \bibfield  [0]{\@secondoftwo}%
\providecommand \translation [1]{[#1]}%
\providecommand \BibitemOpen [0]{}%
\providecommand \bibitemStop [0]{}%
\providecommand \bibitemNoStop [0]{.\EOS\space}%
\providecommand \EOS [0]{\spacefactor3000\relax}%
\providecommand \BibitemShut  [1]{\csname bibitem#1\endcsname}%
\bibitem [{\citenamefont
  {Klapdor-Kleingrothaus}(2004)}]{KlapdorKleingrothaus:2004wj}%
  \BibitemOpen
  \bibfield  {author} {\bibinfo {author} {\bibfnamefont {H.~V.}\ \bibnamefont
  {Klapdor-Kleingrothaus}},\ }\href@noop {} {\bibfield  {journal} {\bibinfo
  {journal} {Phys. Lett. B},\ }\textbf {\bibinfo {volume} {586}},\ \bibinfo
  {pages} {198} (\bibinfo {year} {2004})},\ \Eprint
  {http://arxiv.org/abs/hep-ph/0404088} {hep-ph/0404088} \BibitemShut {NoStop}%
\bibitem [{\citenamefont {Avignone{III}}\ \emph {et~al.}(2008)\citenamefont
  {Avignone{III}}, \citenamefont {Elliott},\ and\ \citenamefont
  {Engel}}]{Avignone:2007fu}%
  \BibitemOpen
  \bibfield  {author} {\bibinfo {author} {\bibfnamefont {F.~T.}\ \bibnamefont
  {Avignone{III}}}, \bibinfo {author} {\bibfnamefont {S.~R.}\ \bibnamefont
  {Elliott}}, \ and\ \bibinfo {author} {\bibfnamefont {J.}~\bibnamefont
  {Engel}},\ }\Doi {10.1103/RevModPhys.80.481} {\bibfield  {journal} {\bibinfo
  {journal} {Rev. Mod. Phys.},\ }\textbf {\bibinfo {volume} {80}},\ \bibinfo
  {eid} {481} (\bibinfo {year} {2008})},\ \bibinfo {note} {arXiv:0708.1033
  [nucl-ex]}\BibitemShut {NoStop}%
\bibitem [{\citenamefont {Kortelainen}\ and\ \citenamefont
  {Suhonen}(2007){\natexlab{a}}}]{Kortelainen:2007rh}%
  \BibitemOpen
  \bibfield  {author} {\bibinfo {author} {\bibfnamefont {M.}~\bibnamefont
  {Kortelainen}}\ and\ \bibinfo {author} {\bibfnamefont {J.}~\bibnamefont
  {Suhonen}},\ }\href@noop {} {\bibfield  {journal} {\bibinfo  {journal} {Phys.
  Rev. C},\ }\textbf {\bibinfo {volume} {75}},\ \bibinfo {pages} {51303}
  (\bibinfo {year} {2007}{\natexlab{a}})},\ \Eprint
  {http://arxiv.org/abs/arXiv:0705.0469 [nucl-th]} {arXiv:0705.0469 [nucl-th]}
  \BibitemShut {NoStop}%
\bibitem [{\citenamefont {Kortelainen}\ and\ \citenamefont
  {Suhonen}(2007){\natexlab{b}}}]{Kortelainen:2007mn}%
  \BibitemOpen
  \bibfield  {author} {\bibinfo {author} {\bibfnamefont {M.}~\bibnamefont
  {Kortelainen}}\ and\ \bibinfo {author} {\bibfnamefont {J.}~\bibnamefont
  {Suhonen}},\ }\href@noop {} {\bibfield  {journal} {\bibinfo  {journal} {Phys.
  Rev. C},\ }\textbf {\bibinfo {volume} {76}},\ \bibinfo {pages} {24315}
  (\bibinfo {year} {2007}{\natexlab{b}})},\ \Eprint
  {http://arxiv.org/abs/arXiv:0708.0115 [nucl-th]} {arXiv:0708.0115 [nucl-th]}
  \BibitemShut {NoStop}%
\bibitem [{\citenamefont {Rodin}\ \emph {et~al.}(2007)\citenamefont {Rodin},
  \citenamefont {Faessler}, \citenamefont {\v{S}imkovic},\ and\ \citenamefont
  {Vogel}}]{Rodin:2007fz}%
  \BibitemOpen
  \bibfield  {author} {\bibinfo {author} {\bibfnamefont {V.~A.}\ \bibnamefont
  {Rodin}}, \bibinfo {author} {\bibfnamefont {A.}~\bibnamefont {Faessler}},
  \bibinfo {author} {\bibfnamefont {F.}~\bibnamefont {\v{S}imkovic}}, \ and\
  \bibinfo {author} {\bibfnamefont {P.}~\bibnamefont {Vogel}},\ }\href@noop {}
  {\bibfield  {journal} {\bibinfo  {journal} {Nucl. Phys. A},\ }\textbf
  {\bibinfo {volume} {793}},\ \bibinfo {pages} {213} (\bibinfo {year}
  {2007})},\ \Eprint {http://arxiv.org/abs/arXiv:0706.4304 [nucl-th]}
  {arXiv:0706.4304 [nucl-th]} \BibitemShut {NoStop}%
\bibitem [{\citenamefont {Caurier}\ \emph
  {et~al.}(2008){\natexlab{a}}\citenamefont {Caurier}, \citenamefont
  {Men\'{e}ndez}, \citenamefont {Nowacki},\ and\ \citenamefont
  {Poves}}]{Caurier:2007wq}%
  \BibitemOpen
  \bibfield  {author} {\bibinfo {author} {\bibfnamefont {E.}~\bibnamefont
  {Caurier}}, \bibinfo {author} {\bibfnamefont {J.}~\bibnamefont
  {Men\'{e}ndez}}, \bibinfo {author} {\bibfnamefont {F.}~\bibnamefont
  {Nowacki}}, \ and\ \bibinfo {author} {\bibfnamefont {A.}~\bibnamefont
  {Poves}},\ }\Doi {10.1103/PhysRevLett.100.052503} {\bibfield  {journal}
  {\bibinfo  {journal} {Phys. Rev. Lett.},\ }\textbf {\bibinfo {volume}
  {100}},\ \bibinfo {pages} {52503} (\bibinfo {year} {2008}{\natexlab{a}})},\
  \Eprint {http://arxiv.org/abs/0709.2137} {arXiv:0709.2137 [nucl-th]}
  \BibitemShut {NoStop}%
\bibitem [{\citenamefont {Barea}\ and\ \citenamefont
  {Iachello}(2009)}]{Barea:2009ia}%
  \BibitemOpen
  \bibfield  {author} {\bibinfo {author} {\bibfnamefont {J.}~\bibnamefont
  {Barea}}\ and\ \bibinfo {author} {\bibfnamefont {F.}~\bibnamefont
  {Iachello}},\ }\Doi {10.1103/PhysRevC.79.044301} {\bibfield  {journal}
  {\bibinfo  {journal} {Phys. Rev. C},\ }\textbf {\bibinfo {volume} {79}},\
  \bibinfo {pages} {044301} (\bibinfo {year} {2009})}\BibitemShut {NoStop}%
\bibitem [{\citenamefont {Chandra}\ \emph {et~al.}(2009)\citenamefont
  {Chandra}, \citenamefont {Chaturvedi}, \citenamefont {Rath}, \citenamefont
  {Raina},\ and\ \citenamefont {Hirsch}}]{Chandra:2009my}%
  \BibitemOpen
  \bibfield  {author} {\bibinfo {author} {\bibfnamefont {R.}~\bibnamefont
  {Chandra}}, \bibinfo {author} {\bibfnamefont {K.}~\bibnamefont {Chaturvedi}},
  \bibinfo {author} {\bibfnamefont {P.~K.}\ \bibnamefont {Rath}}, \bibinfo
  {author} {\bibfnamefont {P.~K.}\ \bibnamefont {Raina}}, \ and\ \bibinfo
  {author} {\bibfnamefont {J.~G.}\ \bibnamefont {Hirsch}},\ }\href@noop {}
  {\bibfield  {journal} {\bibinfo  {journal} {Eur. Phys. Lett. A},\ }\textbf
  {\bibinfo {volume} {86}},\ \bibinfo {pages} {32001} (\bibinfo {year}
  {2009})}\BibitemShut {NoStop}%
\bibitem [{\citenamefont {Doi}\ \emph {et~al.}(1985)\citenamefont {Doi},
  \citenamefont {Kotani},\ and\ \citenamefont {Takasugi}}]{Doi:1985dx}%
  \BibitemOpen
  \bibfield  {author} {\bibinfo {author} {\bibfnamefont {M.}~\bibnamefont
  {Doi}}, \bibinfo {author} {\bibfnamefont {T.}~\bibnamefont {Kotani}}, \ and\
  \bibinfo {author} {\bibfnamefont {E.}~\bibnamefont {Takasugi}},\ }\href@noop
  {} {\bibfield  {journal} {\bibinfo  {journal} {Prog. Theor. Phys. Suppl.},\
  }\textbf {\bibinfo {volume} {83}},\ \bibinfo {pages} {1} (\bibinfo {year}
  {1985})}\BibitemShut {NoStop}%
\bibitem [{\citenamefont {Simkovic}\ \emph {et~al.}(2004)\citenamefont
  {Simkovic}, \citenamefont {Pacearescu},\ and\ \citenamefont
  {Faessler}}]{Simkovic:2004pf}%
  \BibitemOpen
  \bibfield  {author} {\bibinfo {author} {\bibfnamefont {F.}~\bibnamefont
  {Simkovic}}, \bibinfo {author} {\bibfnamefont {L.}~\bibnamefont
  {Pacearescu}}, \ and\ \bibinfo {author} {\bibfnamefont {A.}~\bibnamefont
  {Faessler}},\ }\href@noop {} {\bibfield  {journal} {\bibinfo  {journal}
  {Nucl. Phys. A},\ }\textbf {\bibinfo {volume} {733}},\ \bibinfo {pages} {321}
  (\bibinfo {year} {2004})}\BibitemShut {NoStop}%
\bibitem [{\citenamefont {\'{A}lvarez Rodr\'{\i}guez}\ \emph
  {et~al.}(2004)\citenamefont {\'{A}lvarez Rodr\'{\i}guez}, \citenamefont
  {Sarriguren}, \citenamefont {Moya~de Guerra}, \citenamefont {Pacearescu},
  \citenamefont {Faessler},\ and\ \citenamefont
  {\v{S}imkovic}}]{AlvarezRodriguez:2004uk}%
  \BibitemOpen
  \bibfield  {author} {\bibinfo {author} {\bibfnamefont {R.}~\bibnamefont
  {\'{A}lvarez Rodr\'{\i}guez}}, \bibinfo {author} {\bibfnamefont
  {P.}~\bibnamefont {Sarriguren}}, \bibinfo {author} {\bibfnamefont
  {E.}~\bibnamefont {Moya~de Guerra}}, \bibinfo {author} {\bibfnamefont
  {L.}~\bibnamefont {Pacearescu}}, \bibinfo {author} {\bibfnamefont
  {A.}~\bibnamefont {Faessler}}, \ and\ \bibinfo {author} {\bibfnamefont
  {F.}~\bibnamefont {\v{S}imkovic}},\ }\Doi {10.1103/PhysRevC.70.064309}
  {\bibfield  {journal} {\bibinfo  {journal} {Phys. Rev.},\ }\textbf {\bibinfo
  {volume} {C70}},\ \bibinfo {pages} {64309} (\bibinfo {year} {2004})},\
  \Eprint {http://arxiv.org/abs/nucl-th/0411039} {arXiv:nucl-th/0411039}
  \BibitemShut {NoStop}%
\bibitem [{\citenamefont {Caurier}\ \emph
  {et~al.}(2008){\natexlab{b}}\citenamefont {Caurier}, \citenamefont
  {Nowacki},\ and\ \citenamefont {Poves}}]{Caurier:2007qn}%
  \BibitemOpen
  \bibfield  {author} {\bibinfo {author} {\bibfnamefont {E.}~\bibnamefont
  {Caurier}}, \bibinfo {author} {\bibfnamefont {F.}~\bibnamefont {Nowacki}}, \
  and\ \bibinfo {author} {\bibfnamefont {A.}~\bibnamefont {Poves}},\
  }\href@noop {} {\bibfield  {journal} {\bibinfo  {journal} {Eur. Phys. J.
  A.},\ }\textbf {\bibinfo {volume} {36}},\ \bibinfo {pages} {195} (\bibinfo
  {year} {2008}{\natexlab{b}})},\ \Eprint {http://arxiv.org/abs/arXiv:0709.0277
  [nucl-th]} {arXiv:0709.0277 [nucl-th]} \BibitemShut {NoStop}%
\bibitem [{\citenamefont {Simkovic}(2007)}]{Simkovic:2007me}%
  \BibitemOpen
  \bibfield  {author} {\bibinfo {author} {\bibfnamefont {F.}~\bibnamefont
  {Simkovic}},\ }in\ \href@noop {} {\emph {\bibinfo {booktitle} {Proceedings of
  the Workshop on the Calculation of Double Beta Decay Matrix Elements,
  MEDEX07}}},\ Vol.\ \bibinfo {volume} {942},\ \bibinfo {editor} {edited by\
  \bibinfo {editor} {\bibfnamefont {O.}~\bibnamefont {Civitarese}}, \bibinfo
  {editor} {\bibfnamefont {I.}~\bibnamefont {Stekl}}, \ and\ \bibinfo {editor}
  {\bibfnamefont {J.}~\bibnamefont {Suhonen}}}\ (\bibinfo  {publisher} {AIP
  Conf. Proc.},\ \bibinfo {year} {2007})\BibitemShut {NoStop}%
\bibitem [{\citenamefont {Miller}\ and\ \citenamefont
  {Spencer}(1976)}]{Miller:1975hu}%
  \BibitemOpen
  \bibfield  {author} {\bibinfo {author} {\bibfnamefont {G.~A.}\ \bibnamefont
  {Miller}}\ and\ \bibinfo {author} {\bibfnamefont {J.~E.}\ \bibnamefont
  {Spencer}},\ }\href@noop {} {\bibfield  {journal} {\bibinfo  {journal} {Ann.
  Phys.},\ }\textbf {\bibinfo {volume} {100}},\ \bibinfo {pages} {562}
  (\bibinfo {year} {1976})}\BibitemShut {NoStop}%
\bibitem [{\citenamefont {Engel}\ and\ \citenamefont
  {Hagen}(2009)}]{Engel:2009ha}%
  \BibitemOpen
  \bibfield  {author} {\bibinfo {author} {\bibfnamefont {J.}~\bibnamefont
  {Engel}}\ and\ \bibinfo {author} {\bibfnamefont {G.}~\bibnamefont {Hagen}},\
  }\href@noop {} {\bibfield  {journal} {\bibinfo  {journal} {Phys. Rev. C},\
  }\textbf {\bibinfo {volume} {79}},\ \bibinfo {pages} {064317} (\bibinfo
  {year} {2009})},\ \Eprint {http://arxiv.org/abs/0904.1709} {arXiv:0904.1709
  [nucl-th]} \BibitemShut {NoStop}%
\bibitem [{\citenamefont {Simkovic}\ \emph {et~al.}(2009)\citenamefont
  {Simkovic}, \citenamefont {Faessler}, \citenamefont {Muther}, \citenamefont
  {Rodin},\ and\ \citenamefont {Stauf}}]{Simkovic:2009pp}%
  \BibitemOpen
  \bibfield  {author} {\bibinfo {author} {\bibfnamefont {F.}~\bibnamefont
  {Simkovic}}, \bibinfo {author} {\bibfnamefont {A.}~\bibnamefont {Faessler}},
  \bibinfo {author} {\bibfnamefont {H.}~\bibnamefont {Muther}}, \bibinfo
  {author} {\bibfnamefont {V.}~\bibnamefont {Rodin}}, \ and\ \bibinfo {author}
  {\bibfnamefont {M.}~\bibnamefont {Stauf}},\ }\href@noop {} {\bibfield
  {journal} {\bibinfo  {journal} {Phys. Rev. C},\ }\textbf {\bibinfo {volume}
  {79}},\ \bibinfo {pages} {055501} (\bibinfo {year} {2009})},\ \Eprint
  {http://arxiv.org/abs/0902.0331} {arXiv:0902.0331 [nucl-th]} \BibitemShut
  {NoStop}%
\bibitem [{\citenamefont {Men\'{e}ndez}\ \emph {et~al.}(2009)\citenamefont
  {Men\'{e}ndez}, \citenamefont {Poves}, \citenamefont {Caurier},\ and\
  \citenamefont {Nowacki}}]{Menendez:2009def}%
  \BibitemOpen
  \bibfield  {author} {\bibinfo {author} {\bibfnamefont {J.}~\bibnamefont
  {Men\'{e}ndez}}, \bibinfo {author} {\bibfnamefont {A.}~\bibnamefont {Poves}},
  \bibinfo {author} {\bibfnamefont {E.}~\bibnamefont {Caurier}}, \ and\
  \bibinfo {author} {\bibfnamefont {F.}~\bibnamefont {Nowacki}},\ }in\
  \href@noop {} {\emph {\bibinfo {booktitle} {Proceedings of the International
  School of Physics Enrico Fermi, Measurements of Neutrino Mass}}},\ \bibinfo
  {editor} {edited by\ \bibinfo {editor} {\bibfnamefont {F.}~\bibnamefont
  {Ferroni}, \bibfnamefont {F.~Vissani}}\ and\ \bibinfo {editor} {\bibfnamefont
  {C.}~\bibnamefont {Brofferio}}}\ (\bibinfo  {publisher} {IOS Press},\
  \bibinfo {year} {2009})\ \bibinfo {note} {arXiv:0809.2183}\BibitemShut
  {NoStop}%
\bibitem [{\citenamefont {\v{S}imkovic}\ \emph {et~al.}(2008)\citenamefont
  {\v{S}imkovic}, \citenamefont {Faessler}, \citenamefont {Rodin},
  \citenamefont {Vogel},\ and\ \citenamefont {Engel}}]{Simkovic:2007vu}%
  \BibitemOpen
  \bibfield  {author} {\bibinfo {author} {\bibfnamefont {F.}~\bibnamefont
  {\v{S}imkovic}}, \bibinfo {author} {\bibfnamefont {A.}~\bibnamefont
  {Faessler}}, \bibinfo {author} {\bibfnamefont {V.~A.}\ \bibnamefont {Rodin}},
  \bibinfo {author} {\bibfnamefont {P.}~\bibnamefont {Vogel}}, \ and\ \bibinfo
  {author} {\bibfnamefont {J.}~\bibnamefont {Engel}},\ }\Doi
  {10.1103/PhysRevC.77.045503} {\bibfield  {journal} {\bibinfo  {journal}
  {Phys. Rev. C},\ }\textbf {\bibinfo {volume} {77}},\ \bibinfo {pages} {45503}
  (\bibinfo {year} {2008})},\ \Eprint {http://arxiv.org/abs/0710.2055}
  {arXiv:0710.2055 [nucl-th]} \BibitemShut {NoStop}%
\bibitem [{\citenamefont {Feldmeier}\ \emph {et~al.}(1998)\citenamefont
  {Feldmeier}, \citenamefont {Neff}, \citenamefont {Roth},\ and\ \citenamefont
  {Schnack}}]{Feldmeier:1997zh}%
  \BibitemOpen
  \bibfield  {author} {\bibinfo {author} {\bibfnamefont {H.}~\bibnamefont
  {Feldmeier}}, \bibinfo {author} {\bibfnamefont {T.}~\bibnamefont {Neff}},
  \bibinfo {author} {\bibfnamefont {R.}~\bibnamefont {Roth}}, \ and\ \bibinfo
  {author} {\bibfnamefont {J.}~\bibnamefont {Schnack}},\ }\href@noop {}
  {\bibfield  {journal} {\bibinfo  {journal} {Nucl. Phys. A},\ }\textbf
  {\bibinfo {volume} {632}},\ \bibinfo {pages} {61} (\bibinfo {year} {1998})},\
  \Eprint {http://arxiv.org/abs/nucl-th/9709038} {nucl-th/9709038} \BibitemShut
  {NoStop}%
\bibitem [{\citenamefont {Escuderos}\ \emph {et~al.}(2010)\citenamefont
  {Escuderos}, \citenamefont {Faessler}, \citenamefont {Rodin},\ and\
  \citenamefont {Simkovic}}]{Escuderos:2010xx}%
  \BibitemOpen
  \bibfield  {author} {\bibinfo {author} {\bibfnamefont {A.}~\bibnamefont
  {Escuderos}}, \bibinfo {author} {\bibfnamefont {A.}~\bibnamefont {Faessler}},
  \bibinfo {author} {\bibfnamefont {V.}~\bibnamefont {Rodin}}, \ and\ \bibinfo
  {author} {\bibfnamefont {F.}~\bibnamefont {Simkovic}},\ }\href@noop {} {
  (\bibinfo {year} {2010})},\ \Eprint {http://arxiv.org/abs/arXiv:1001.3519v2
  [nucl-th]} {arXiv:1001.3519v2 [nucl-th]} \BibitemShut {NoStop}%
\bibitem [{\citenamefont {Engel}\ \emph {et~al.}(1989)\citenamefont {Engel},
  \citenamefont {Vogel}, \citenamefont {Ji},\ and\ \citenamefont
  {Pittel}}]{Engel:1989vj}%
  \BibitemOpen
  \bibfield  {author} {\bibinfo {author} {\bibfnamefont {J.}~\bibnamefont
  {Engel}}, \bibinfo {author} {\bibfnamefont {P.}~\bibnamefont {Vogel}},
  \bibinfo {author} {\bibfnamefont {X.}~\bibnamefont {Ji}}, \ and\ \bibinfo
  {author} {\bibfnamefont {S.}~\bibnamefont {Pittel}},\ }\href@noop {}
  {\bibfield  {journal} {\bibinfo  {journal} {Phys. Lett. B},\ }\textbf
  {\bibinfo {volume} {225}},\ \bibinfo {pages} {5} (\bibinfo {year}
  {1989})}\BibitemShut {NoStop}%
\end{thebibliography}%

\end{document}